# Detection of Influenza A Virus Nucleoprotein Through the Self-Assembly of Nanoparticles in Magnetic Particle Spectroscopy-Based Bioassays: A Method for Rapid, Sensitive, and Wash-free Magnetic Immunoassays

*Kai Wu, Jinming Liu, Renata Saha, Diqing Su, Venkatramana D. Krishna, Maxim C-J Cheeran\* and Jian-Ping Wang\**

Dr. K. Wu, J. Liu, R. Saha, Prof. J.-P. Wang
Department of Electrical and Computer Engineering
University of Minnesota, Minneapolis, MN 55455, United States
Email: jpwang@umn.edu
D. Su
Department of Chemical Engineering and Material Science
University of Minnesota, Minneapolis, MN 55455, United States
Dr. V.D. Krishna, Prof. M.C-J. Cheeran
Department of Veterinary Population Medicine
University of Minnesota, St. Paul, Minnesota 55108, USA
E-mail: cheeran@umn.edu

**Abstract**

Magnetic nanoparticles (MNPs) with proper surface functionalization have been extensively applied as labels for magnetic immunoassays, carriers for controlled drug/gene delivery, tracers and contrasts for magnetic imaging, etc. Here, we introduce a new biosensing scheme based on magnetic particle spectroscopy (MPS) and the self-assembly of MNPs to quantitatively detect H1N1 nucleoprotein molecules. MPS monitors the harmonics of oscillating MNPs as a metric for the freedom of rotational motion, thus indicating the bound states of MNPs. These harmonics can be readily collected from nanogram quantities of iron oxide nanoparticles within 10 s. H1N1 nucleoprotein molecule hosts multiple different epitopes that forms binding sites for many IgG polyclonal antibodies. Anchoring IgG polyclonal antibodies onto MNPs triggers the cross-linking between MNPs and H1N1 nucleoprotein molecules, thereby forming MNP self-assemblies. Using MPS and the self-assembly of MNPs, we achieved the sensitivity of 44 nM (442 pmole) for detecting H1N1 nucleoprotein. In addition, the morphologies and the hydrodynamic sizes of the MNP self-assemblies are characterized to verify the MPS results. Different MNP self-assembly models such as classical cluster, open ring tetramer, chain model as well as multimers (from dimer to pentamer) are proposed in this paper. Herein, we claim the feasibility of using MPS and the self-assembly of MNPs as a new biosensing scheme for detecting ultralow concentrations of target biomolecules, which can be employed as rapid, sensitive, and wash-free magnetic immunoassays.

## 1. Introduction

Influenza viruses belong to the family Orthomyxoviridae and contain single stranded, negative sense, segmented RNA genome. Based on their differences in nucleoprotein (NP) and matrix (M) protein, influenza viruses are classified into four types, A, B, C, and D.[1–4] Influenza A viruses (IAV) are further sub-divided based on combinations of their surface glycoproteins hemagglutinin (HA) and neuraminidase (NA) into different subtypes. There are 18 different types of HA (H1 to H18) and 11 different types of NA (N1 to N11). Each IAV subtype is virus with one type of H and N combination (such as the H1N1 subtype investigated in this work). IAV infects many vertebrates including humans and is responsible for seasonal epidemics of acute respiratory illness known as influenza or flu. IAV poses a significant public health concern that causes substantial morbidity and mortality and has the ability to cause worldwide pandemics.[5,6] According to World Health Organization (WHO), influenza viruses are responsible for 250,000 to 500,000 deaths annually. Rapid, accurate and sensitive method for early diagnosis of IAV infections are critical for rapid initiation of antiviral therapy to control infection and to reduce the impact of possible influenza pandemics.

IAV nucleoprotein (NP) is a basic protein that binds to viral RNA along with polymerase and is the most abundant component of the ribonucleoprotein (RNP) complex.[7] Each RNP consists of one genomic RNA segment associated with a single trimeric polymerase complex and multiple NP monomers.[8,9] In addition to binding single stranded RNA, IAV nucleoprotein has been shown to self-assemble to form large oligomeric complexes.[10] Moreover, nucleoprotein is well conserved among different IAV strains isolated from different host species with amino acid difference of less than 11%.[11] These properties make IAV nucleoprotein a good target for detection of multiple IAV subtypes.

The conventional detection schemes of IAV can be divided into several categories based on the type of assays employed. Immunofluorescence assays use either direct or indirect fluorescent antibody staining techniques to detect IAV. The signal can be readout through fluorescent microscopes with sensitivities of 60 – 80%. However, they are unable to distinguish different IAV subtypes.[12] Serological tests such as hemagglutination inhibition assays (HAI), microneutralization or virus neutralization assays (VN) and enzyme linked immunosorbent assays (ELISA) are the most commonly used techniques to detect influenza virus-specific antibody responses. But they usually need paired samples with strict requirements of the sample collection time.[13] Nucleic acid-based assays such as reverse transcription-polymerase chain reaction (RT-PCR) are based on the detection of DNA/RNA sequences of the viruses. Apart from the fact that long reaction time (1 – 8 hr) is required, RT-PCR is one of the most expensive testing techniques.[14] To cut down the time of detection, rapid influenza diagnostic tests (RIDTs) have also been developed with RT-PCR and ELISA as the two gold standards.[15,16] Although the total testing time is less than 30 min with high specificities (95 – 99%), the sensitivities of the RIDTs are lower than other techniques (10 – 70%) and false negative results should be considered.[17,18] Giant magnetoresistive (GMR) biosensors have also been reported for the detection of IAV both in lab-based and point-of-care platforms.[19–21] However, this technique is limited by the complexity of nanofabricating the GMR sensors as well as the cost per chip/test.

Magnetic particle spectroscopy (MPS), a novel measurement tool that closely related to magnetic particle imaging (MPI), has emerged recent years as an alternative to the aforementioned techniques and is gaining increasing attentions in the area of volumetric-based bioassays.[22–26] In MPS, also called 0D MPI, oscillating magnetic fields periodically drive magnetic nanoparticles (MNPs) into saturated states where the magnetic responses contain not only the fundamental driving field frequencies but also a series of harmonic frequencies. These harmonics are very useful metrics for characterizing the MNP ferrofluids such as the viscosity and temperature of the solution as well as the conjugations of chemical compounds on the MNPs.[26–31] Nowadays, with the ease of fabricating and surface-functionalizing MNPs, MNPs functionalized with ligands, DNA, RNA, and protein molecules can be readily applied as probes, labels, contrasts, and tracers for different bioassays.[26,31–39] Unlike the surface-based biosensing platforms (ELISA, GMR, Hall effect sensors, etc.) that directly detect individual target objects near the sensing elements, the volumetric-based biosensing platforms measure the analytical signals that directly come from the entire detection volume, making the bioassays simple and fast.[40–43] Representative examples of volumetric-based sensors are the conventional superconducting quantum interference devices (SQUIDs), the nuclear magnetic resonance (NMR) devices, and the magnetic susceptometers.[44–50] MPS is one type of volumetric-based biosensing platform where the MNPs are acting as the mini-sensing probes and their dynamic magnetic responses are used as metrics for the characterization of target analytes from the fluidic samples.[51–55]

In this current work, we report the MPS-based bioassay platform as a rapid, sensitive, and wash-free method for the detection of H1N1 nucleoprotein. Due to the IgG antibodies anchored on the MNPs, cross-linking takes place between MNPs and H1N1 nucleoprotein which forms MNP clusters. Characterization of these MNP clusters reveal their controlled formation and hence in the upcoming sections we have referred to them as MNP 'self-assemblies'. The 3$^{rd}$ and the 5$^{th}$ harmonics along with the 3$^{rd}$ over the 5$^{th}$ harmonic ratios (R35) can be used as metrics for the characterization of target analyte concentrations from fluidic samples. These harmonics can be detected from nanogram quantities of iron oxide MNPs within 10 s. We show that H1N1 nucleoprotein can be detected with high sensitivity (44 nM or 442 pmole) using this detection scheme. By combining MPS technology and the self-assembly of MNPs, we are able to achieve rapid, sensitive, and wash-free magnetic bioassays. Furthermore, this detection scheme utilizing the self-assembly of MNPs is suitable for detecting and quantifying a wide range of biomarkers/analytes.

## 2. Results and Discussion

### 2.1 Conjugating IgG Polyclonal Antibodies on MNPs.

To immobilize rabbit IgG polyclonal antibodies onto MNPs, the zero-length carbodiimide crosslinker EDC is used to couple the carboxyl groups from MNPs to the primary amines from IgG. As shown in Figure 1(a):i – iii, the EDC reacts with carboxylic acid group from MNP to form an O-acylisourea active ester that can be easily displaced by nucleophilic attack from primary amino group on IgG. In this step, 1.5 mL microcentrifuge tube containing 200 μL, 0.29 nmole/mL SHP-25 MNP suspension is mixed with 200 μL, 25 mM MES and 10 μL, 10 mg/mL EDC, react at room temperature for 15 min with continuous mixing. Then, 100 μL, 5 μg/mL polyclonal IgG is added to the suspension and react at room temperature for 2.5 hours with continuous mixing. As is shown in Figure 1(a):iii – v, the primary amine from IgG forms an amide bond with the original carboxyl group and an EDC by-product is released as a soluble urea derivative (not shown in the figure). Hereafter, 100 μL, 100 mM Trizma hydrochloride buffer is added to the suspension and react at room temperature for 30 minutes with continuous mixing to quench the EDC activation reaction. The MNP and IgG antibody complexes (denoted as 'MNP+Aby' in this paper) are formed and the unconjugated IgG polyclonal antibodies are removed by ultra-centrifuging the MNP ferrofluid at 11 000 RPM, acceleration 11 200 g, for 45 min (PowerSpin™ BX Centrifuge), then the supernatant is removed and the remaining MNP+Aby complexes are resuspended in PBS to a total volume of 200 μL (in order to maintain the original concentration of 0.29 nmole/mL nanoparticles). This wash step is repeated for three times to thoroughly remove the unconjugated IgG polyclonal antibodies. Using the same method, eight samples each containing 100 μL, 0.29 nmole/mL MNP+Aby complexes in PBS buffer are prepared (labeled as samples I – VIII). The as-prepared MNPs anchored with rabbit IgG polyclonal antibodies are ready to be used for bioassays and can be stored at 2°C – 8°C under sterile condition for one month without detectable loss of magnetic properties of MNPs and activity of IgG. This test kit containing microliter volume surface functionalized MNPs allows for future on field bioassays along with a portable MPS platform.

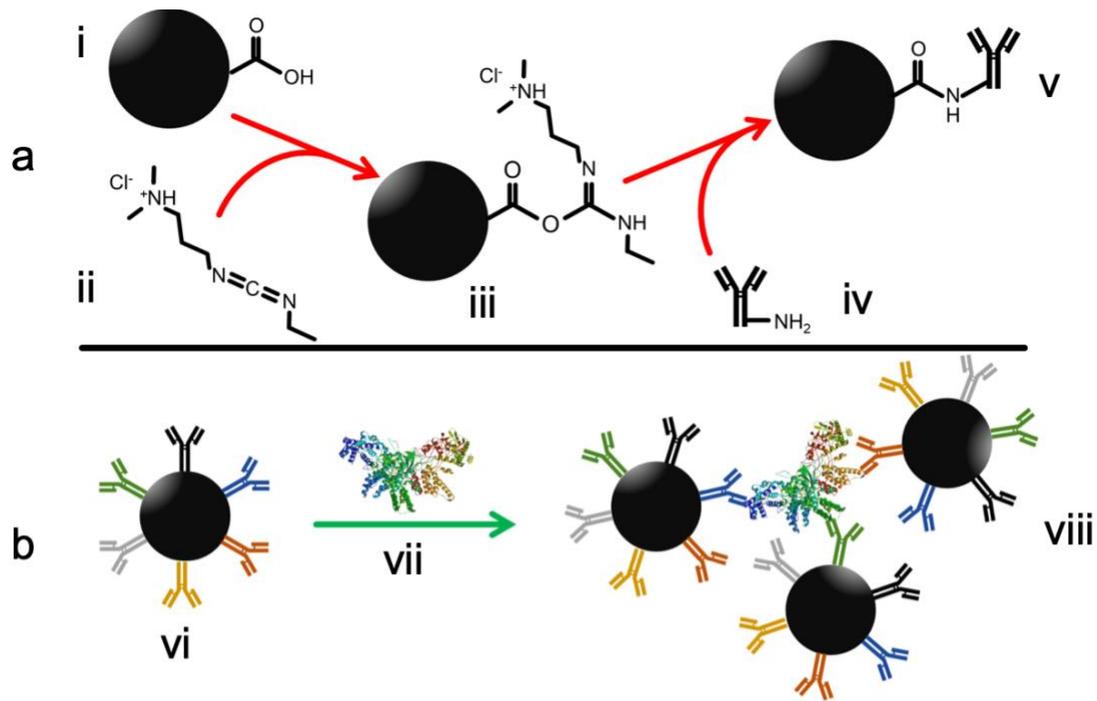
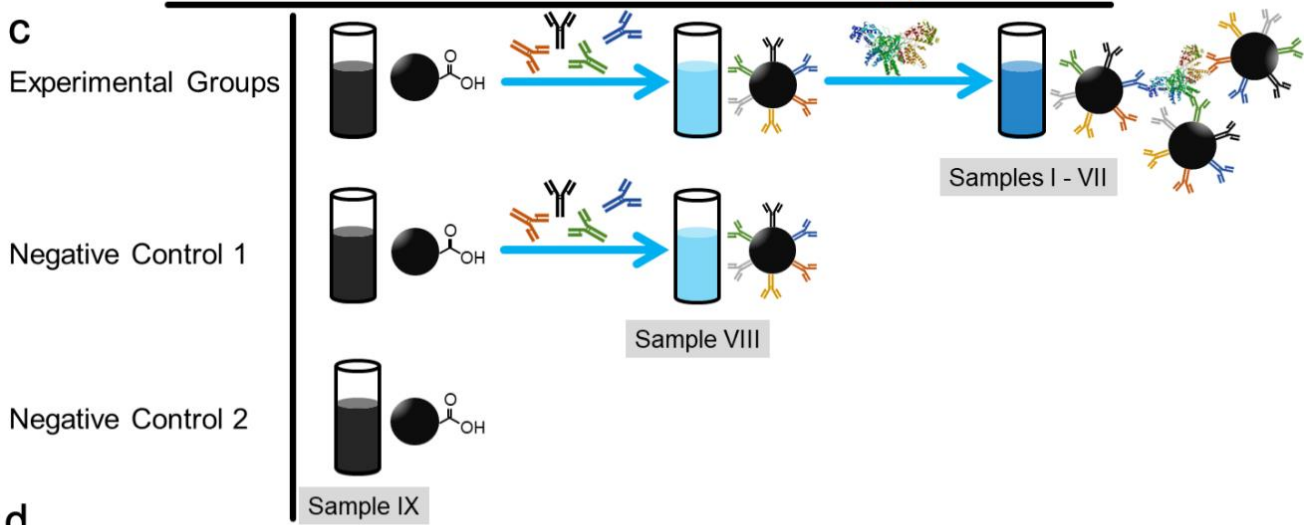
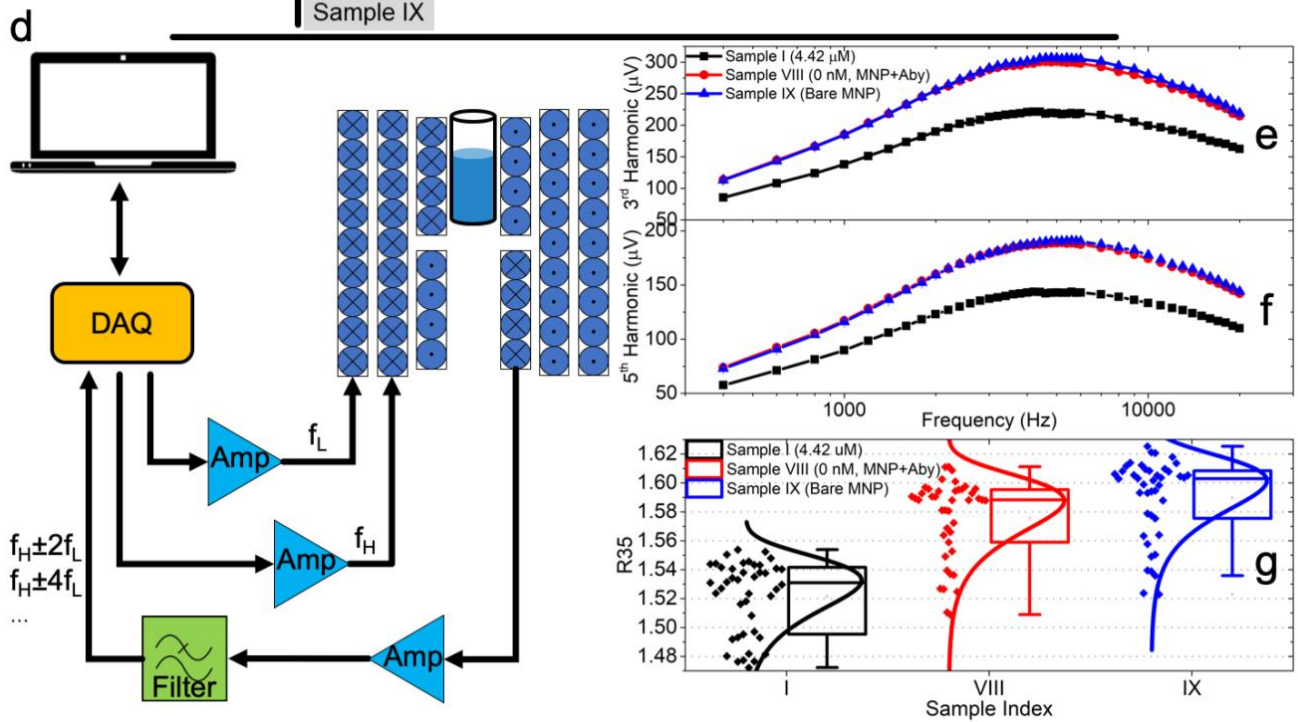

**Figure 1. (a)** EDC (carbodiimide) crosslinking reaction scheme. i) SHP-25 MNP with carboxylic acid group; ii) EDC crosslinker; iii) O-acylisourea active ester (unstable) on MNP; iv) The polyclonal IgG with primary amine group; v) The reaction of O-acylistourea with the amine from IgG forms stable amide bond on MNP. **(b)** Schematic view of the self-assembly of MNPs through proper choice of IgG polyclonal antibodies and target antigen hosting multiple different epitopes. vi) SHP-25 MNP surface conjugated with IgG polyclonal antibodies; vii) H1N1 nucleoprotein (AAM75159.1) (Met1-Gly490) possessing multiple different epitopes that allows a number of IgG polyclonal antibodies to bind; viii) The final product of the self-assembled MNPs. **(c)** Sample preparation flow charts of the experimental and negative control groups. Nine MNP samples are prepared: sample index I – VII are MNP-antibody complexes in the presence of different concentrations of H1N1 nucleoprotein; sample index VIII is MNP-antibody complexes in the absence of H1N1 nucleoprotein (denoted as '0 nM (MNP+Aby)' in this paper); sample index IX is bare MNP suspension (denoted as 'Bare MNP' in this paper). The details of samples I – IX are listed in **Table 1**. **(d)** Schematic view of MPS system setups. **(e) and (f)** are the $3^{rd}$ and the $5^{th}$ harmonics along varying driving field frequencies (only samples I, VIII, and IX are plotted) collected by the MPS system. **(g)** Boxplots of the harmonic ratios (R35) collected from samples I, VIII, and IX. Note that figures **(a) – (c)** are not drawn to scale.

## 2.2 The Experimental and Control Groups.

As shown in Figure 1(c), to the as-prepared MNP-IgG samples I – VII were added 100 μL H1N1 nucleoprotein of varying concentrations (from 4.42 μM down to 44 nM as listed in Table 1). To sample VIII, 100 μL PBS buffer was added (instead of antigen) as negative control group #1 (denoted as '0 nM (MNP+Aby)' in this paper). Sample IX containing 100 μL, 0.29 nmole/mL SHP-25 MNP and 100 μL PBS buffer (without antibody or H1N1 nucleoprotein) is prepared as negative control group #2 (denoted as 'Bare MNP' in this paper). Samples I – IX are incubated at room temperature for 2 hours with continuous mixing to allow the cross-linking of MNPs and H1N1 nucleoprotein molecules, followed by storage at 2°C – 8°C under sterile condition before the MPS measurements.

Table 1. Experimental design of samples I - IX

| Sample Index | | I | II | III | IV | V | VI | VII | VIII | IX |
|---|---|---|---|---|---|---|---|---|---|---|
| **MNP** | | 100 μL | 100 μL | 100 μL | 100 μL | 100 μL | 100 μL | 100 μL | 100 μL | 100 μL |
| **Polyclonal IgG** | | Yes | Yes | Yes | Yes | Yes | Yes | Yes | Yes | No |
| | Concentration | 4.42 μM | 2.21 μM | 884 nM | 442 nM | 221 nM | 88 nM | 44 nM | 0 nM | 0 nM |

| H1N1 Nucleoprotein | Volume* | 100 µL | 100 µL | 100 µL | 100 µL | 100 µL | 100 µL | 100 µL | 100 µL | 100 µL |

*Note: to samples I – VII we added 100 µL of varying concentrations of H1N1 nucleoprotein and to samples VIII & IX are added with 100 µL PBS buffer.

**2.3 Self-Assembly of MNPs.**

Different degrees of MNP self-assemblies (here different degrees indicate the number and average sizes of MNP self-assemblies) are formed in samples I – VII due to the fact that the H1N1 nucleoprotein molecule has multiple epitopes recognized by polyclonal antibodies (see Figure 1(b)). As a result, each H1N1 nucleoprotein is bound to more than one MNPs. On the other hand, each MNP functionalized with multiple IgG polyclonal antibodies is able to bind with multiple H1N1 nucleoprotein molecules. Thus, the cross-linking of MNPs and H1N1 nucleoproteins leads to different degrees of MNP self-assemblies depending on the number/concentration of H1N1 nucleoproteins in the MNP ferrofluid.

**2.4 Detection of H1N1 Nucleoprotein *via* MPS and Magnetic Relaxation Dynamics.**

When the MNPs are suspended in a solution under oscillating magnetic fields, there are two mechanisms governing the rotation of magnetic moments in response to the magnetic fields (see Scheme 1(a) & (b)). One is the intrinsic Néel motion (rotating magnetic moment inside the stationary MNP) and the other is the extrinsic Brownian motion (rotating the entire MNP along with its magnetic moment). In principle, the joint effects of Néel and Brownian relaxation mechanisms contribute to the macroscopic magnetic response of MNP ferrofluid sample subjected to external oscillating magnetic fields. Herein, we use the SHP-25 MNPs with averaged core diameter of 25 nm, where Brownian motion is the dominating relaxation mechanism and thus determines the macroscopic magnetic responses of MNPs (in this case the effect of Néel motion is negligible, see Scheme 1(c)). For these Brownian motion-dominated MNPs, their macroscopic magnetic responses are directly related to the overall hydrodynamic sizes.

MPS-based bioassay uses the harmonics of oscillating MNPs as a metric for the freedom of rotational motion, thus indicating the degree of MNP self-assembly. In the presence of H1N1 nucleoprotein molecules, the polyclonal IgG functionalized MNPs can bind to one nucleoprotein molecule or cross-link to multiple nucleoprotein molecules, thus, forming MNP self-assemblies (see Scheme 1(d)). Such an interaction leads to an increased hydrodynamic size of the MNPs, resulting in an increased Brownian relaxation time, which is monitored by the MPS. The 3$^{rd}$ and the 5$^{th}$ harmonics are immediately recorded at $f_H \pm 2f_L$ and $f_H \pm 2f_L$. In addition, the ratio of the 3$^{rd}$ over the 5$^{th}$ harmonic (R35) is used as a MNP quantity-independent metric for the detection of H1N1 nucleoprotein. Herein, we report using the 3$^{rd}$ and the 5$^{th}$ harmonics along with the 3$^{rd}$ over the 5$^{th}$ harmonic ratio (R35) as metrics for the characterization of target analyte concentrations from fluidic samples (see example plots in Figure 1(e) – (g)).

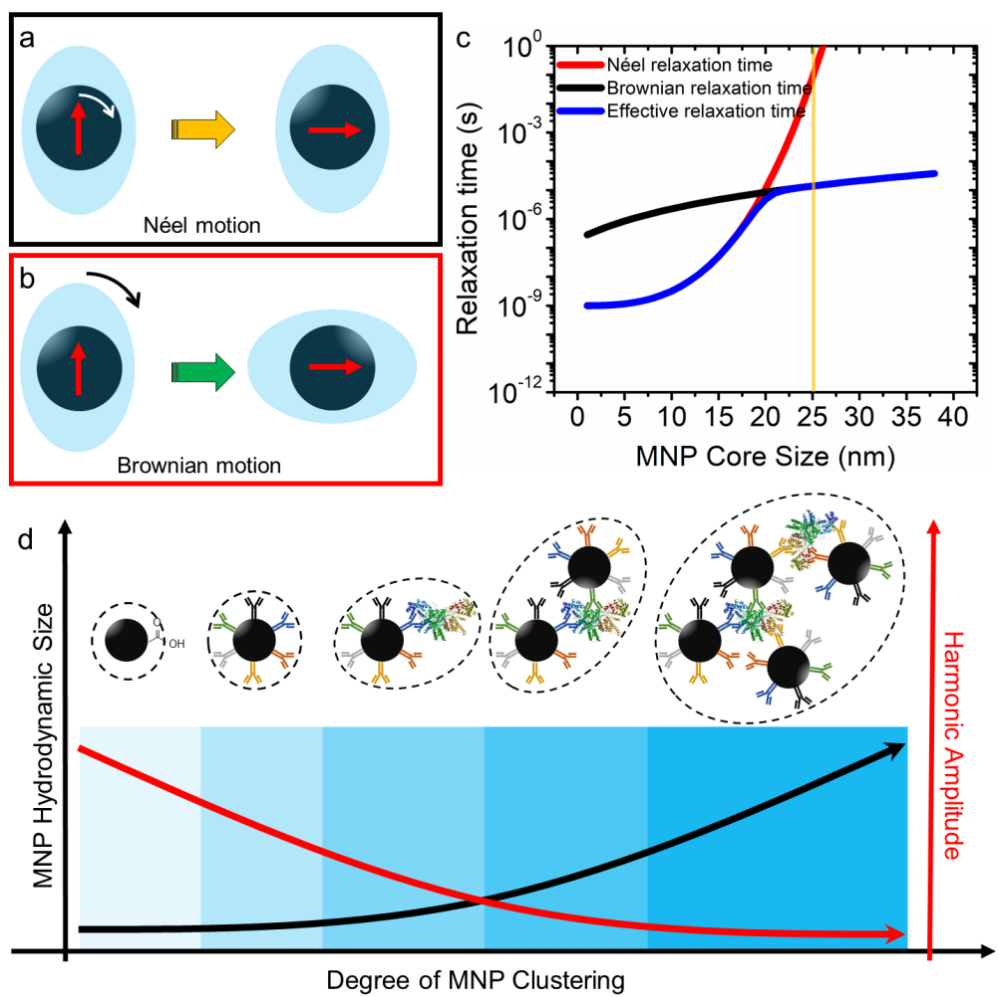

**Scheme 1. (a) Néel motion is the rotation of magnetic moment inside the stationary MNP. (b) Brownian motion is the rotation of the entire MNP along with its magnetic moment. (c) The Néel, Brownian, and effective relaxation times are depicted as a function of iron oxide MNP core size, assuming MNPs are dispersed in a solution with viscosity η=1 cP and temperature T=300 K. The vertical orange line highlights that the Brownian motion is the dominating relaxation mechanism for iron oxide MNPs with core size of 25 nm. The magnetic relaxation time models can be found from Supporting Information Note S2. (d) As the degree of MNP self-assembly increases in the presence of H1N1 nucleoprotein, the average hydrodynamic sizes of particles (MNPs and MNP self-assemblies) increases, as a result, the measured harmonic amplitude decreases. The dashed black lines represent the hydrodynamic sizes of MNPs and MNP self-assemblies. Note that figures (a), (b) and (d) are schematics and are not drawn to scale.**

## 2.5 The 3rd and the 5th Harmonics as Metrics for the Detection of H1N1 Nucleoprotein.

The magnetic responses of MNPs from the as-prepared samples I – IX are monitored and recorded by our lab-based MPS system. One negative control group 1 (sample VIII, see Figure 1(c)) is added to this experiment to validate the selectivity in detection of target biomarker, i.e., H1N1 nucleoprotein. The other negative control group 2 (sample IX, see Figure 1(c)) is added to verify the efficiency of anchoring rabbit IgG polyclonal antibodies on MNPs.

The description of our lab-based MPS system setup can be found in Supporting Information Note S1 and a representative signal chain is shown in Figure 1(d). The frequency $f_H$ of the high-frequency driving field is swept from 400 Hz to 20 kHz, and the frequency of the low-frequency driving field is set at 10 Hz. The amplitudes of the high- and low-frequency driving fields are set at 17 Oe and 170 Oe, respectively. The sampling rate is set at 500 kHz. In each MPS measurement, the background noise floor is collected for 5 s followed by inserting the sample for another 5 s of data collection. The 3rd and the 5th harmonics from the magnetic responses of MNPs are reconstructed from the signal and background noise.

As shown in Figure 2, according to the Faraday's Law of Induction and the Langevin model (see Notes S3 - S6 from Supporting Information), the amplitudes of the 3rd and the 5th harmonics are modulated by the driving field frequency $f_H$. With increasing concentration/quantity of H1N1 nucleoprotein in the samples, MNPs are likely form larger self-assemblies, as a result, the Brownian relaxation time (as well as the effective relaxation time) increases. This increased relaxation time causes an increased phase lag (see Notes S4 from Supporting Information) between the magnetic moments of MNPs and the external oscillating magnetic fields.

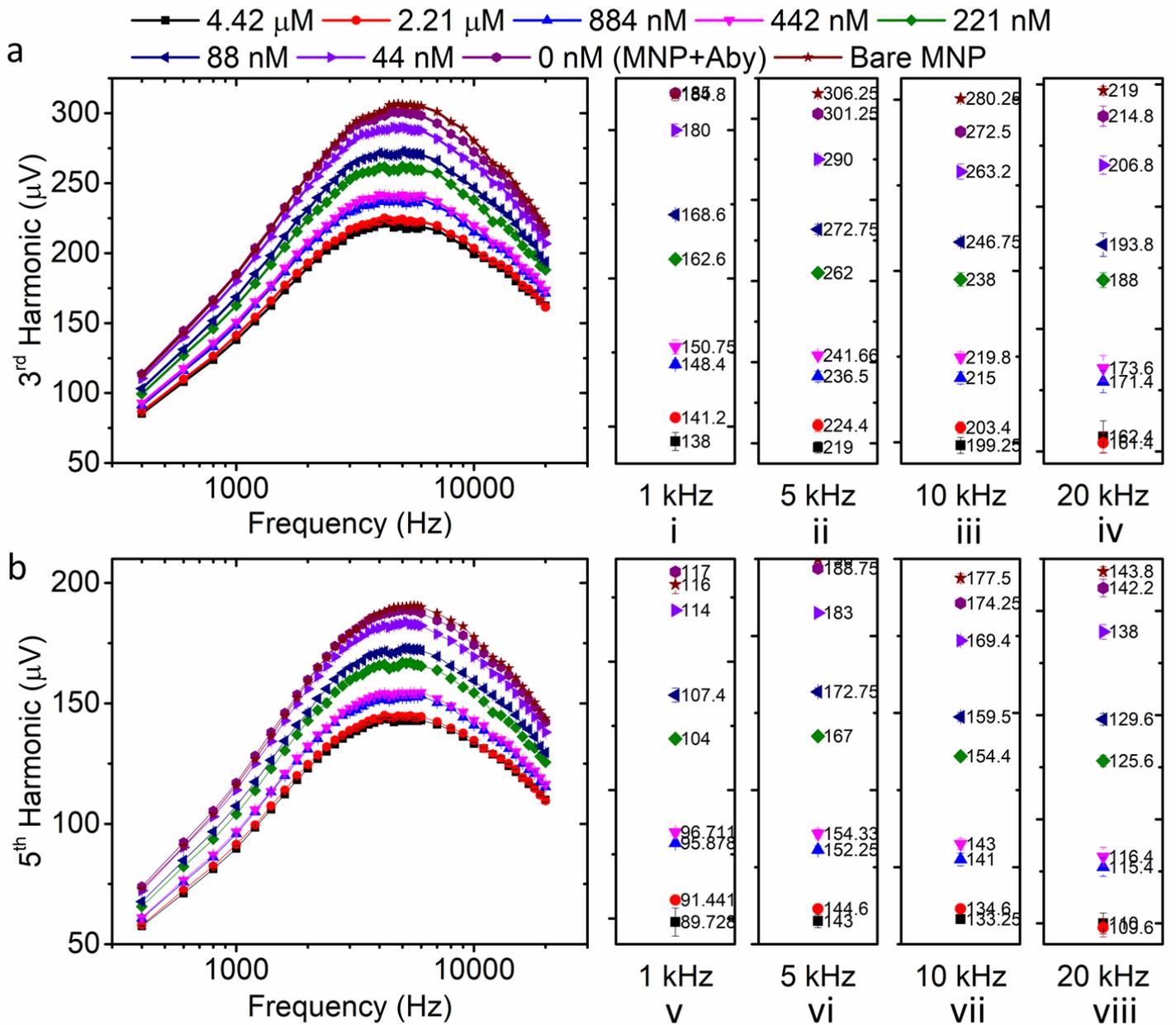

**Figure 2. (a, b)** MPS measurements of the 3rd and the 5th harmonics from samples I – IX at varying driving field frequencies from 400 Hz to 20 kHz. The insets in i) – iv) highlight the 3rd harmonic amplitudes measured at 1 kHz, 5 kHz, 10 kHz and 20 kHz, respectively. Insets v) – viii) highlight the 5th harmonic amplitudes measured at 1 kHz, 5 kHz, 10 kHz and 20 kHz, respectively. The error bar represents standard deviation.

The amplitudes of the 3rd and the 5th harmonics from samples I – IX are summarized in Figure 2(a) & (b), which shows a clear trend that at all driving field frequencies, the samples with higher concentrations of H1N1 nucleoprotein, yield weaker magnetic responses (i.e., smaller harmonic amplitudes). This is due to the following facts: 1) larger MNP self-assemblies are formed in the presence of higher concentration/quantity of H1N1 nucleoprotein; 2) the Brownian motions of MNPs that are cross-linked in the self-assemblies are blocked (i.e., larger hydrodynamic size as shown in Scheme 1(d)); 3) the magnetic moments of these 'trapped' MNPs fail to align with the external magnetic fields, thus, weaker dynamic magnetic responses are observed from these samples. The measured harmonic amplitudes arranged from lowest to highest are: I (4.42 µM) < II (2.21 µM) < III (884 nM) < IV (442 nM) < V (221 nM) < VI (88 nM) < VII (44 nM) < VIII (0 nM, MNP+Aby) < IX (0 nM, bare MNP). Figure 2(a):i – iv and Figure 2(a):v – viii highlight the amplitudes of the $3^{rd}$ and the $5^{th}$ harmonics at driving field frequencies of 1 kHz, 5 kHz, 10 kHz, and 20 kHz, respectively. These values are summarized in Tables S1 & S2 from Supporting Information.

The differences in the harmonics from samples VIII and IX indicate that the rabbit IgG polyclonal antibodies have been successfully anchored onto MNPs. The conjugation of antibodies onto MNPs could slightly increase the hydrodynamic size of MNPs (see Scheme 1(d)), which is also evident in the DLS results in Figure 5(a) – (e) where Figure 5(d) & (e) are the averaged hydrodynamic sizes are 48.4 nm and 46.3 nm for negative control groups, VIII (MNP+Aby) and IX (bare MNP), respectively.

**2.6 Harmonic Ratio (R35) as MNP Quantity-Independent Metric for the Detection of H1N1 Nucleoprotein.**

As the harmonics are largely dependent on the number of MNPs from the sample, the MPS results could be biased by the deviations of MNP quantities from one sample to another, especially for the detection of very low concentration target biomarkers. Besides the metrics of the $3^{rd}$ and the $5^{th}$ harmonics for the characterization of H1N1 nucleoproteins. The harmonic ratio, R35, has also been proposed as MNP quantity-independent metric for the detection of biomarkers (see Notes S7 from Supporting Information).[26,31,56] Herein, the harmonic ratios R35 from samples I – IX at various driving field frequencies are summarized in Figure 3(a). Again, a clear trend is noticed that at all driving field frequencies, the samples with higher concentrations of H1N1 nucleoprotein yield smaller R35 values. Figure 3(a):i – iv highlights the R35 values from each sample at driving field frequencies of 1 kHz, 5 kHz, 10 kHz, and 20 kHz, respectively.

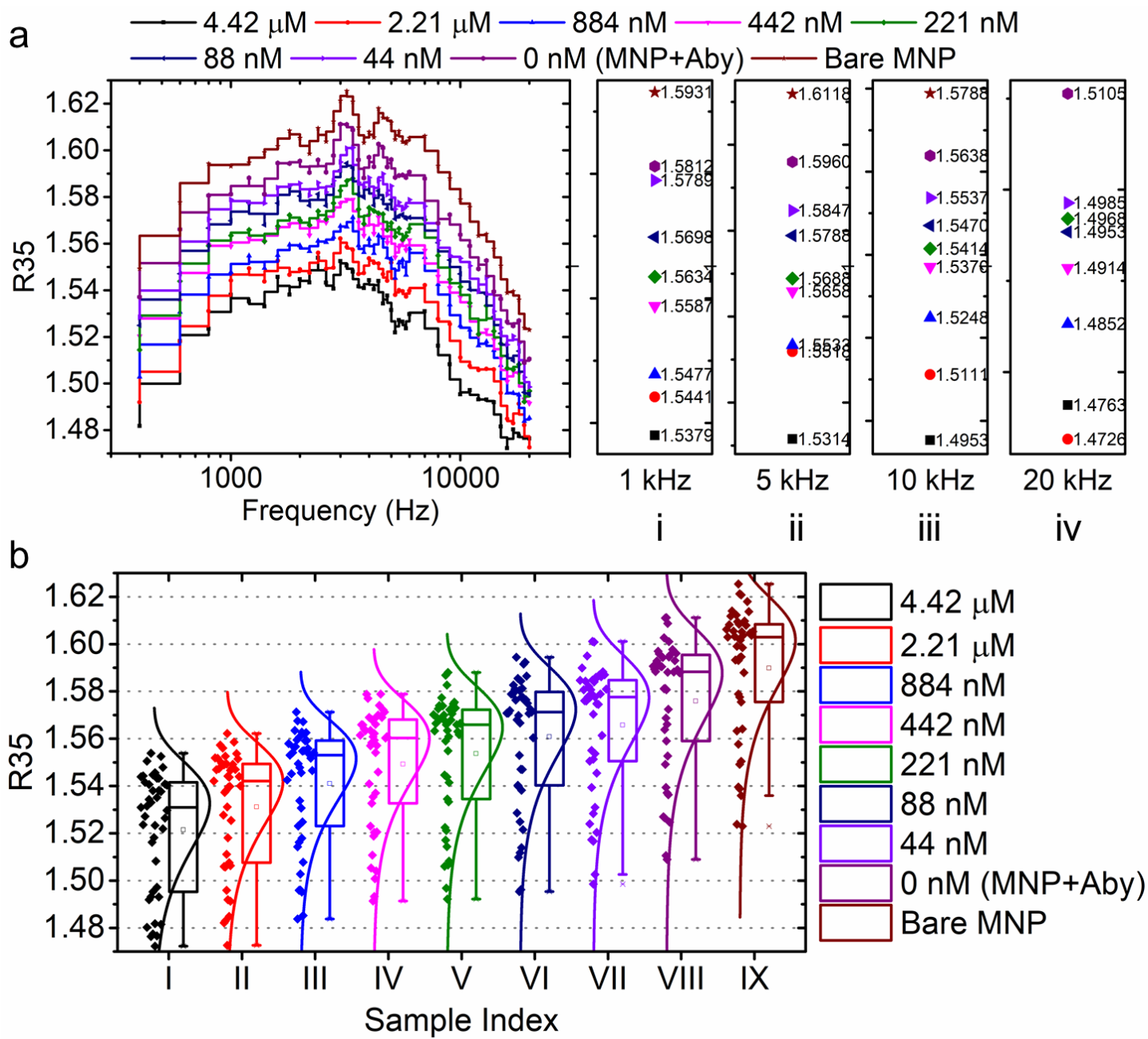

**Figure 3.** MPS measurements of the ratio of the 3$^{rd}$ over the 5$^{th}$ harmonics (R35) from samples I – IX. **(a)** Harmonic ratios, R35, from samples I – IX as we vary the driving field frequencies from 400 Hz to 20 kHz. Insets i) – iv) highlight the R35 measured at 1 kHz, 5 kHz, 10 kHz and 20 kHz, respectively. **(b)** Boxplot of R35 from samples I – IX. The first quartile (Q1, 25$^{th}$ percentile), median (Q2, 50$^{th}$ percentile), and third quartile (Q3, 75$^{th}$ percentile) values from samples I – IX are summarized in **Table 2**. R35 heatmap of samples I – IX is reported in **Notes S8 from Supporting Information**.

Figure 3(b) is the boxplot of R35 values from MNP samples at all driving field frequencies. The statistical first quartile (Q1), median (Q2), third quartile (Q3) R35 values from samples I – IX are summarized in Table 2. Again, there is a clear trend that the R35 value decreases as the concentration of H1N1 nucleoprotein increases.

Table 2. The 25$^{th}$, 50$^{th}$, and 75$^{th}$ percentiles R35 values of 9 samples in the boxplot from Figure 3(b)

| Sample Index | I | II | III | IV | V | VI | VII | VIII | IX |
|---|---|---|---|---|---|---|---|---|---|
| 25$^{th}$ Percentile (Q1) | 1.541 | 1.549 | 1.559 | 1.568 | 1.573 | 1.580 | 1.585 | 1.595 | 1.610 |
| 50$^{th}$ Percentile (Q2) | 1.531 | 1.542 | 1.553 | 1.560 | 1.566 | 1.571 | 1.578 | 1.588 | 1.603 |
| 75$^{th}$ Percentile (Q3) | 1.495 | 1.508 | 1.523 | 1.533 | 1.534 | 1.540 | 1.550 | 1.559 | 1.576 |

## 2.7 Sensitivity of H1N1 Nucleoprotein Detection.

The MPS responses of MNPs versus H1N1 nucleoprotein concentrations ranging from 44 nM to 4.42 µM (4.4 – 442 pmole) are investigated in Figure 4. Overall, the harmonic amplitudes and the R35 values increase as the concentration of H1N1 nucleoprotein decreases. It should be noted that both the 3$^{rd}$ and the 5$^{th}$ harmonics show significantly weaker capabilities in detecting samples with low concentrations of H1N1 nucleoproteins (see Notes S9 from Supporting Information). Furthermore, neither the 3$^{rd}$ harmonic nor the 5$^{th}$ harmonic succeed in distinguishing samples I (4.42 µM) and II (2.21 µM) at driving field frequency of 20 kHz, where the concentrations of H1N1 nucleoprotein from these samples are extremely high. On the other hand, the harmonic ratio, R35, demonstrates comparable capabilities in distinguishing high and low concentrations of H1N1 nucleoprotein samples (see Figure S3(c) from Supporting Information). However, it fails to detect the differences in samples I (4.42 µM) and II (2.21 µM) at 20 kHz driving field frequency. All of the metrics reported in this paper, the 3$^{rd}$ harmonic, the 5$^{th}$ harmonic, and the harmonic ratio R35, achieved a detection limit of 44 nM (442 pmole) for H1N1 nucleoprotein.

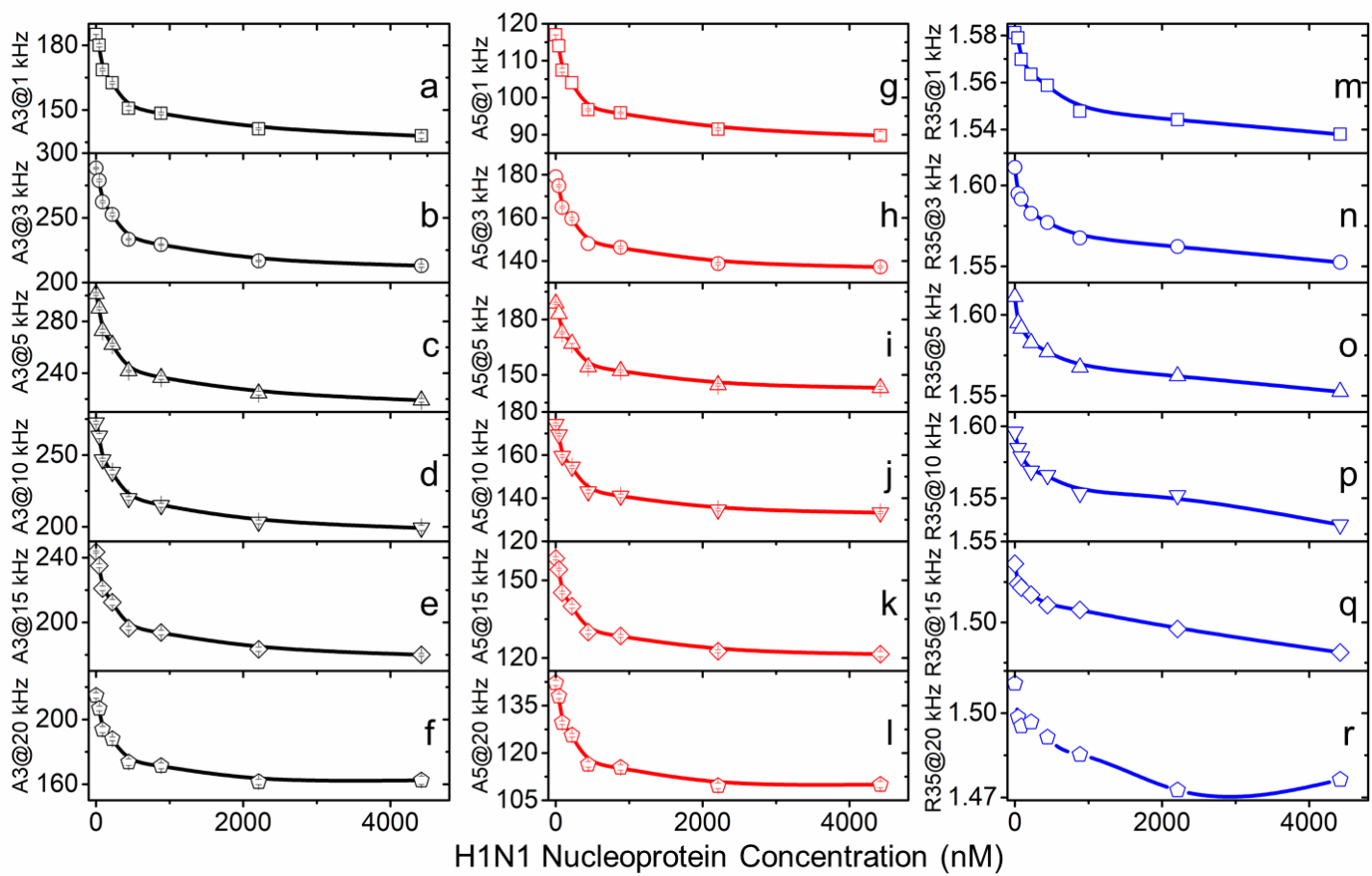

**Figure 4.** (a) – (f) are the amplitudes of the 3$^{rd}$ harmonics from samples I – VIII at driving field frequencies of (a) 1 kHz; (b) 3 kHz; (c) 5 kHz; (d) 10 kHz; (e) 15 kHz; and (f) 20 kHz, respectively. (g) – (l) are the amplitudes of the 5$^{th}$ harmonics from samples I – VIII at driving field frequencies of (g) 1 kHz; (h) 3 kHz; (i) 5 kHz; (j) 10 kHz; (k) 15 kHz; and (l) 20 kHz, respectively. (m) – (r) are the harmonic ratios, R35, from samples I – VIII at driving field frequencies of (m) 1 kHz; (n) 3 kHz; (o) 5 kHz; (p) 10 kHz; (q) 15 kHz; and (r) 20 kHz, respectively. The error bar represents standard deviation.

**2.8 Hydrodynamic Size Analysis on the MNP Self-Assemblies.**

The hydrodynamic sizes of the samples are characterized by Dynamic Light Scatter (DLS, Microtrac NanoFlex). Figure 5(a) – (f) shows the hydrodynamic size distributions of samples II (2.21 µM), IV (442 nM), VI (88 nM), VIII (0 nM, MNP+Aby) and IX (0 nM, bare MNP) with mean values of 58.8 nm, 51.7 nm, 48.7 nm, 48.4 nm and 46.3 nm, respectively. The decreasing mean values from samples II through IX signifies a gradual decrease in the self-assembly degrees of MNPs. Although the primary peaks in the size distributions for all the samples are nearly similar, the main differences come from the peaks at the tails of the size distribution curves (sizes between 200 nm to 300 nm) which is highlighted in the blue area from Figure 5(a) – (e) as well as the black square in Figure 5(f). The tail shows the largest peak in sample II which gradually diminishes in the consecutive samples with little to no peak in the sample IX. The black dashed lines in Figure 5(a) – (e) are the log-normal curve fittings of the particle sizes from samples II, IV, VI, VIII, and IX. Figure 5(f) highlights the differences in the collected DLS size distribution curves from samples II and VIII. Sample II containing 100 µL, 2.21 µM H1N1 nucleoprotein shows the maximum self-assembly degree and hence the largest mean hydrodynamic size is observed. Decreasing concentration of H1N1 nucleoprotein in the remaining samples results in less degrees of MNP self-assembly, hence, smaller mean hydrodynamic size. DLS characteristics of these samples show good agreement with the MPS measurements discussed earlier.

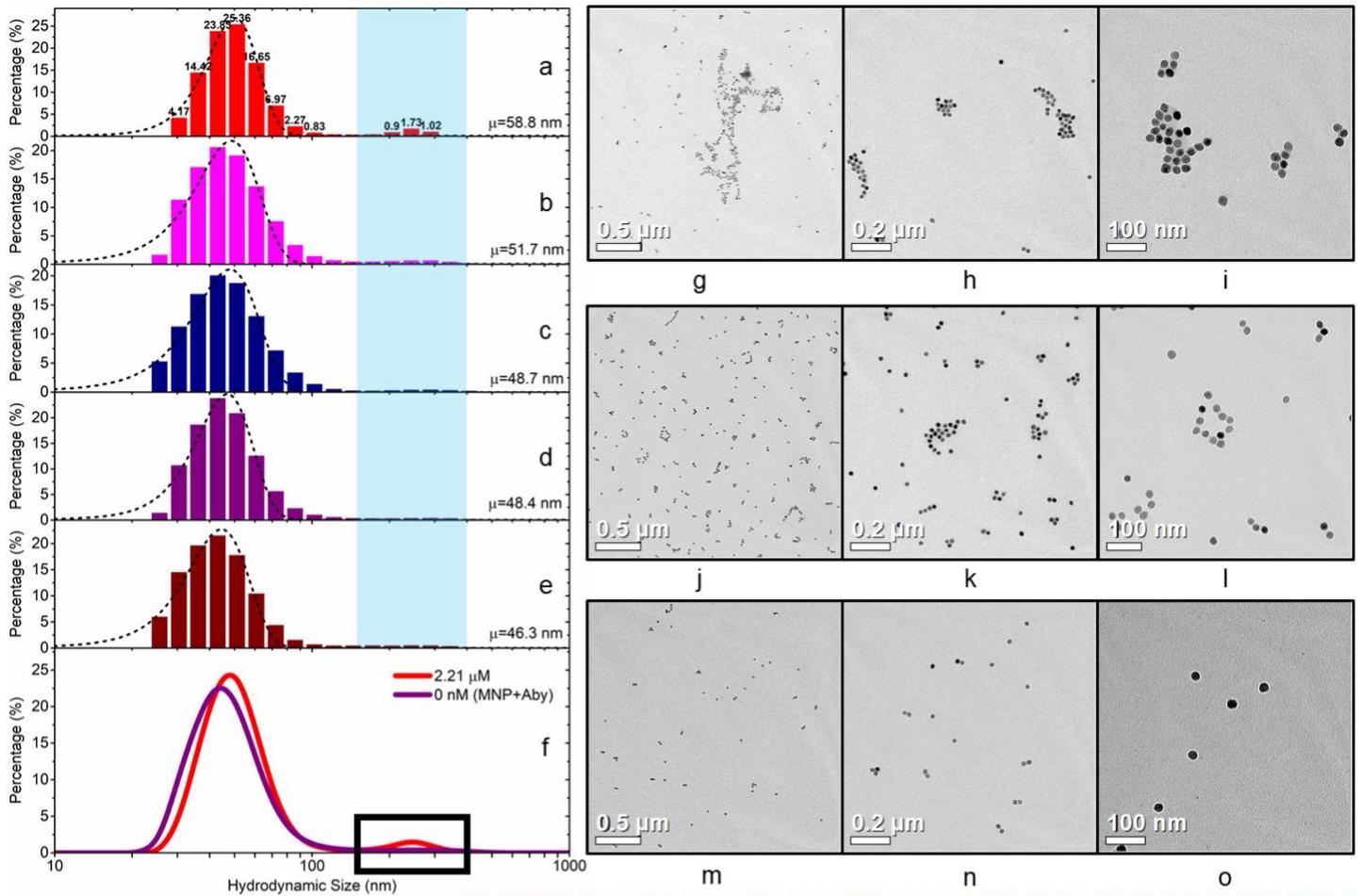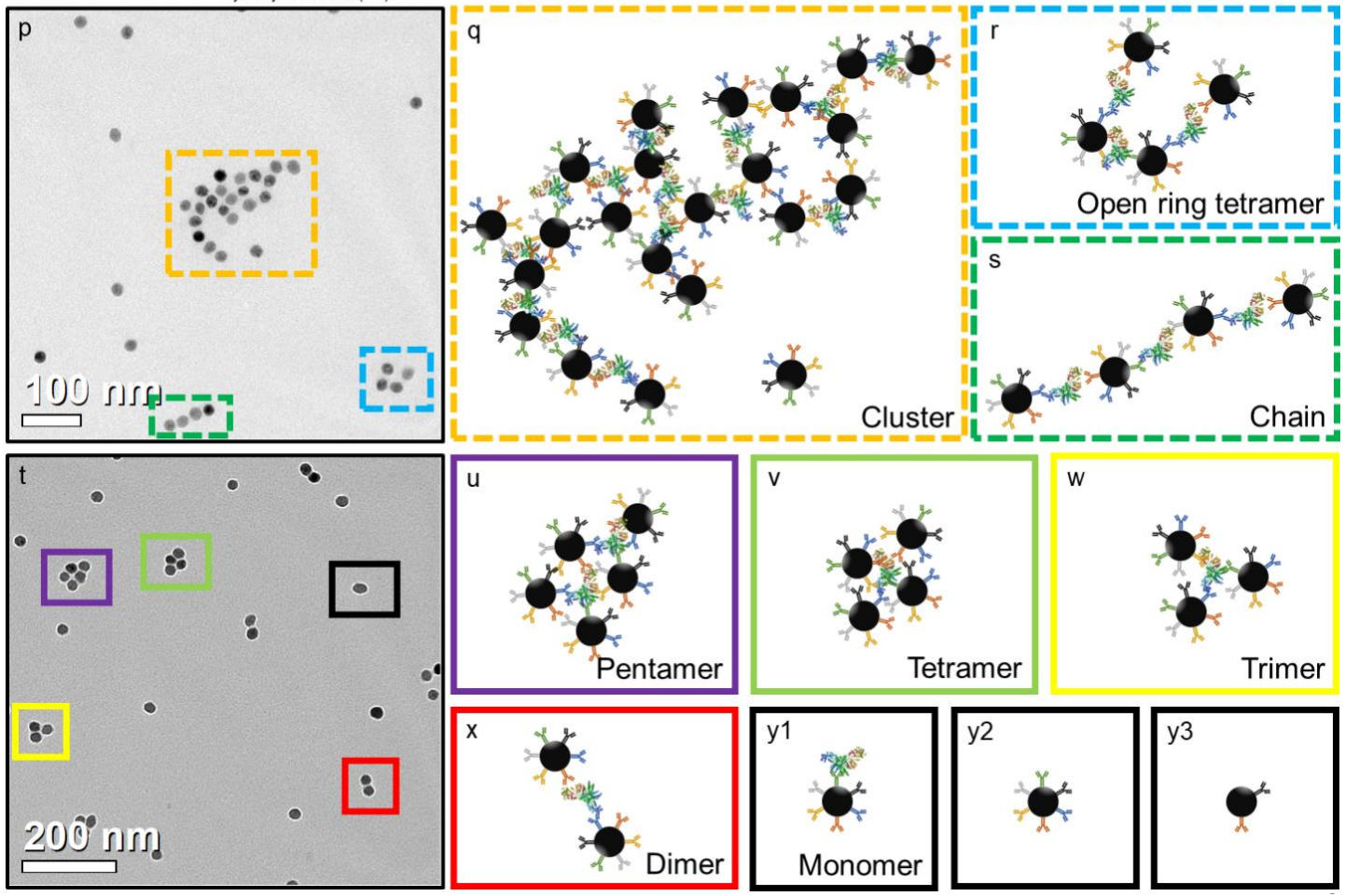

**Figure 5.** Statistical distribution of the hydrodynamic sizes of samples **(a)** II, **(b)** IV, **(c)** VI, **(d)** VIII, and **(e)** IX as characterized by DLS. Dotted lines are the fitted log-normal distribution curves. The μ values represent the statistical mean of the hydrodynamic sizes of the samples. The peaks in the blue highlighted region at the tail show decreasing numbers of MNP self-assemblies from sample II through samples IV, VI, VIII and IX. **(f)** Comparison of the measured DLS size distribution curves between samples II (2.21 µM) and VIII (0 nM, MNP+Aby). The peak at the tail of DLS size distribution curve from sample II is highlighted in the black square. **(g) – (o)** are the bright-field TEM images of samples II **((g) – (i))**, VI **((j) – (l))**, and IX **((m) – (o))**. **(p) and (t)** are the bright-field TEM images from sample VI, highlighting the different shapes of MNP assemblies in the sample. **(q), (r), and (s)** are the MNP self-assembly models: **(q)** the classic MNP self-assembly model (cluster); **(r)** the MNP open ring tetramer model, and **(s)** the MNP chain model representing the orange, blue, and green dotted contours in **(p)**, respectively. **(u), (v), (w), (x), and (y1) – (y3)** are the MNP pentamer, tetramer, trimer, dimer, and monomer models representing the purple, lime, yellow, red, and black solid contours in **(t)**, respectively. Note that figures **(q) – (s)** and **(u) – (y)** are not drawn to scale.

## 2.9 Morphological Characterization on the MNP Self-Assemblies.

The morphologies of the MNPs and MNP self-assemblies from Samples I – IX are characterized by the transmission electron microscopy (TEM, FEI T12 120 kV). Each TEM sample is prepared by dipping a drop of sample fluid onto a TEM grid (TED PELLA, inc.). The droplet is air-dried at room temperature before taking the TEM characterization. As shown in Figure 5(g) – (o), the bright-field TEM images are taken from samples II (2.21 µM), VI (88 nM), and IX (0 nM, bare MNP). MNPs show average magnetic core size of around 25 nm and narrow size distribution, which is critical to obtain consistent and comparable experimental results. Larger MNP self-assemblies are observed from samples with higher H1N1 nucleoprotein concentration like sample II, which is consistent with the DLS results that higher nucleoprotein concentration yields larger average hydrodynamic size.

It has been mentioned earlier that, the H1N1 nucleoprotein molecule has many epitopes that allow the binding of multiple IgG polyclonal antibodies. In addition, the MNPs are surface-functionalized with multiple IgG polyclonal antibodies that would in turn favors the binding to multiple H1N1 nucleoprotein molecules. This cross-linking between MNPs and H1N1 nucleoproteins causes the formation of different degrees of MNP self-assemblies, which has been captured in the TEM images from sample VI as shown in Figure 5(p) & (t) and is schematically drawn in Figure 5(q) – (y3). Figure 5(q), (r) & (s) shows the classic MNP self-assembly (cluster), an open ring tetramer and a chain tetramer highlighted in the dotted-orange, dotted-blue and dotted-green contours in Figure 5(p), respectively. Figure 5(u), (v), (w) & (x) shows the formation of a pentamer, a tetramer, a trimer and a dimer highlighted in solid-purple, solid-lime, solid-yellow and solid-red contours in Figure 5(t), respectively. The solid-black contour in Figure 5(t) is a monomer and all possible schematic representations for the formation of monomers are drawn in Figure 5(y1) – (y3). The possible reasons for monomer MNPs, as shown in TEM images, in the presence of H1N1 nucleoproteins are: 1) the MNP is conjugated with H1N1 nucleoproteins but is not cross-linked to a second MNP (see Figure 5(y1)); 2) the MNP is not bound to any H1N1 nucleoprotein molecules (see Figure 5(y2)); or 3) the MNP does not have enough IgG polyclonal antibodies anchored on its surface decreasing its ability to bind with H1N1 nucleoprotein molecules and assemble into a cluster (see Figure 5(y3)).

## 3. Conclusions

In this paper, we have successfully demonstrated the feasibility of using a MPS system combined with the self-assembly of MNPs for rapid, sensitive, and wash-free detection of H1N1 nucleoprotein. The H1N1 nucleoprotein molecule has multiple epitopes that serve as binding sites for IgG polyclonal antibodies. Thus, each H1N1 nucleoprotein may be bound to more than one MNP, consequently assembling into clusters. In addition, each MNP functionalized with IgG antibodies specific to multiple epitopes of the H1N1 nucleoprotein enabling the particle to bind to multiple H1N1 nucleoprotein molecules. As a result, the cross-linking of MNPs and H1N1 nucleoproteins leads to different degrees of MNP self-assemblies depending on the number/concentration of H1N1 nucleoprotein molecules in the MNP ferrofluid. Noticeable changes in the macroscopic magnetic responses of MNPs are detected by the MPS system when these ferrofluid samples are subjected to external oscillating magnetic fields. Herein, we have reported the $3^{rd}$ harmonic, the $5^{th}$ harmonic, and the harmonic ratio R35 as metrics for the detection of H1N1 nucleoprotein.

The detection scheme of self-assembly of MNPs along with the harmonic metrics used in this present study was able to detect concentrations as low as 44 nM (442 pmole) of H1N1 nucleoprotein. There are several advantages for using the MPS method: 1) *Rapid*: the testing time for detection is within 10 s; 2) *Sensitive*: detection limits are at the pmole level; 3) *Wash-free*, the as-prepared surface functionalized MNPs can be mixed directly with the fluidic sample without further wash steps and the analytical signals are detected directly from the entire sample volume, making the bioassays simple and fast; 4) *Easy-to-use*: measurement could be carried out on minimally processed biological samples by non-technicians with minimum technical training requirements; 5) *Cost effective*: each bioassay only uses nanogram amounts of iron oxide nanoparticles; 6) *Portable:* the entire MPS system could be assembled onto a single PCB board, allowing to configure a hand-held device for field testing. In short, the approach we report here is a rapid and versatile technique that is suitable for the quantitative detection of a wide range of biomarkers/analytes. MPS is a promising, new, and cheap bioassay platform that has the potential to be applied in many biomedical applications including both *in vivo* and *in vitro* diagnostics.

Furthermore, a test kit containing microliter volumes of surface functionalized MNPs along with a point-of-care MPS device opens the future to field-based bioassays in areas of nanomedicine, food safety, agriculture, and veterinary medicine.

## 4. Experimental Section

*Materials:* The recombinant Influenza A virus (A/Puerto Rico/8/34/Mount Sinai (H1N1)) nucleoprotein (AAM75159.1) (Met1-Gly490) is purchased from Sino Biological Inc. (Catalog#11675-V08B). It comprises of 501 amino acids and has a predicted molecular mass of 56.6 kDa. The H1N1 nucleoprotein is provided as lyophilized powder from sterile 60 mM Tris, 500 mM NaCl, pH 7.4 and stored at -20°C under sterile condition before usage. The lyophilized H1N1 nucleoprotein powder is reconstituted in DI water to prepare a stock solution of 0.25 mg/mL (4.42 μM). The Influenza A nucleoprotein specific polyclonal IgG is produced in rabbits immunized with purified, recombinant Influenza A nucleoprotein (Catalog#11675-V08B; AAM75159.1; Met1-Gly490). This rabbit IgG polyclonal antibody purchased from Sino Biological Inc. (Catalog#11675-T62) is provided as liquid solution with measured concentration of 2 mg/mL, stored at -20°C before usage. The rabbit polyclonal IgG is diluted in PBS to a desired concentration of 5 μg/mL. The MNPs with average particle magnetic core size of 25 nm, concentration of 0.29 nmole/mL nanoparticles (5 mg/mL Fe), suspended in DI water containing 0.02% sodium azide, is purchased from Ocean NanoTech Inc. (Catalog#SHP-25). They are water-soluble iron oxide nanoparticles coated with a monolayer of oleic acid and a monolayer of amphiphilic polymer, their reactive group is carboxylic acid and their zeta potential is between -35 mV and -15 mV. The EDC (1-ethyl-3-(3-dimethylaminopropyl)carbodiimide hydrochloride, $C_8H_{17}N_3 \cdot HCl$) is purchased from Thermo Fisher Scientific (Catalog#22980). It is a water-soluble carbodiimide crosslinker that activates the carboxyl groups on MNPs for spontaneous reaction with primary amines from rabbit IgG polyclonal antibodies, enabling stable antibody conjugation. The EDC powder is dissolved in DI water to a desired concentration of 10 mg/mL before immediate use. The MES (2-(N-Morpholino)ethanesulfonic acid, $C_6H_{13}NO_4S$, Prod. No. M3671), Trizma hydrochloride (Tris(hydroxymethyl)aminomethane hydrochloride, $NH_2C(CH_2OH)_3 \cdot HCl$, Prod. No. T5941), and Phosphate Buffered Saline (PBS, Prod. No. 79378) are purchased from Sigma-Aldrich. The MES powder is dissolved in DI water to a desired concentration of 25 mM, pH 6.0. Trizma hydrochloride powder is dissolved in DI water to a desired concentration of 100 mM, pH 7.4.

*Magnetic Particle Spectroscopy (MPS) System Setups:* The lab-based MPS system setups can be found from Supporting Information Note S1 and the signal chain is shown in Figure 1(d). A laptop installed with LabVIEW sends commands to the Data Acquisition Card (DAQ, NI USB-6289) to generate two sinusoidal waves, which are amplified by two Instrument Amplifiers (IA, HP 6824A). These amplified sinusoidal waves are sent to the primary and secondary coils to produce oscillating magnetic fields: one with frequency $f_L = 10\ Hz$ and amplitude $A_L = 170\ Oe$, the other with frequency $f_H$ varying from 400 Hz to 20 kHz and amplitude $A_H = 17\ Oe$. One pair of differently wound pick-up coils (600 windings clock-wise and 600 windings counter-clock-wise) sense the induced magnetic responses from MNPs and send back to DAQ. The response signals at combinatorial frequencies $f_H \pm 2f_L$ (the 3rd harmonics) and $f_H \pm 2f_L$ (the 5th harmonics) are extracted and analyzed.

**Supporting Information**

Magnetic Particle Spectroscopy System Setups; Magnetic Relaxation Time Models; Langevin Model of Magnetic Responses; The Phase Lag Model; The Induced Voltage in Pick-up Coils; The 3rd and the 5th Harmonics; The 3rd over the 5th Harmonic Ratio (R35) as MNP Quantity-Independent Metric; The R35 Heatmap; Capabilities of the 3rd Harmonic, the 5th Harmonic, and the Harmonic Ratio R35 as Metrics for Distinguishing H1N1 Nucleoprotein Samples with High and Low Concentrations; Table S1. The 3rd harmonic amplitudes (µV) from Figure 2(a):i – iv; Table S2. The 5th harmonic amplitudes (µV) from Figure 2(b):v – viii. This material is available in the supporting information.

**Acknowledgements**

This study was financially supported in part by the Institute of Engineering in Medicine of the University of Minnesota through FY18 IEM Seed Grant Funding Program and the College of Veterinary Medicine Emerging, Zoonotic and Infectious Diseases Signature Program, USDA-NIFA SAES General Agriculture research funds (MIN-62-097). Portions of this work were conducted in the Minnesota Nano Center, which is supported by the National Science Foundation through the National Nano Coordinated Infrastructure Network (NNCI) under Award Number ECCS-1542202. Portions of this work were carried out in the Characterization Facility, University of Minnesota, a member of the NSF-funded Materials Research Facilities Network (www.mrfn.org) via the MRSEC program.

**Conflict of Interest**

The authors declare no competing financial interest.

**Keywords**

magnetic particle spectroscopy, influenza a virus, H1N1, nucleoprotein, magnetic nanoparticle, self-assembly, wash-free, bioassay

# Supporting Information

**Detection of Influenza A Virus Nucleoprotein Through the Self-Assembly of Nanoparticles in Magnetic Particle Spectroscopy-Based Bioassays: A Method for Rapid, Sensitive, and Wash-free Magnetic Immunoassays**

*Kai Wu, Jinming Liu, Renata Saha, Diqing Su, Venkatramana D. Krishna, Maxim C-J Cheeran\* and Jian-Ping Wang\**

Dr. K. Wu, J. Liu, R. Saha, Prof. J.-P. Wang
Department of Electrical and Computer Engineering
University of Minnesota, Minneapolis, MN 55455, United States
Email: jpwang@umn.edu
D. Su
Department of Chemical Engineering and Material Science
University of Minnesota, Minneapolis, MN 55455, United States
Dr. V.D. Krishna, Prof. M.C-J. Cheeran
Department of Veterinary Population Medicine
University of Minnesota, St. Paul, Minnesota 55108, USA
E-mail: cheeran@umn.edu

**Note S1. Magnetic Particle Spectroscopy System Setups**

Figure S1 shows the photographs of lab-based MPS system used in this work.

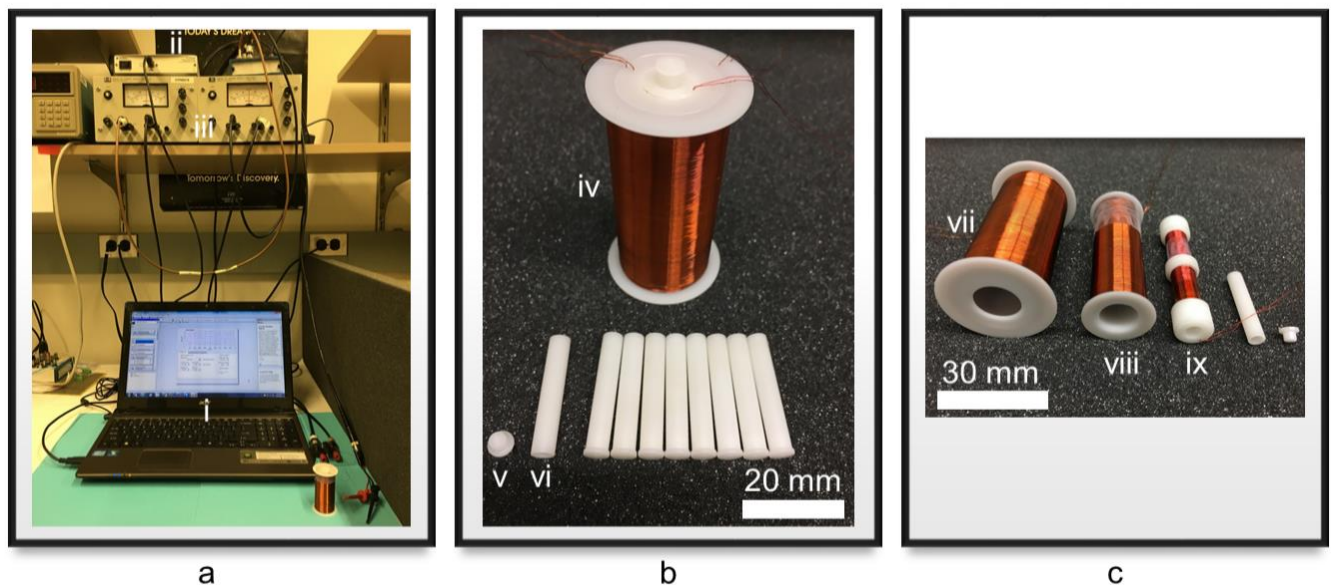

Figure S1. (a) Photograph of lab-based MPS system setups. i) Laptop with LabVIEW that sends commands and receives data; ii) The data acquisition card (DAQ, NI USB-6289); iii) Two instrument amplifiers (IA, HP 6824A). (b) Photograph of one set of coils and the plastic vials with maximum capacity of 300 µL. iv) one set of coils; v) cap of plastic vial; vi) plastic vial. (c) Photograph of primary, secondary, and pick-up coils. vii) primary coil with 1200 windings; viii) Secondary coil with 1200 windings; ix) One pair of pick-up coils with 600 windings clock-wise and 600 windings counter-clock-wise is placed in the center of primary and secondary coils. The plastic vial is inserted into the top half of the pick-up coils (as shown in Figure 1(d)) for the monitoring of MNP magnetic responses. The lower half of pick-up coils are designed to cancel out the voltage caused by the oscillating driving field from primary and secondary coils. As a result, the pick-up coils send back the induced voltage that is only due to the magnetic responses of MNPs.

**Note S2.** Magnetic Relaxation Time Models

When the MNPs are suspended in solution under oscillating magnetic fields, there are two mechanisms governing the rotation of magnetic moments in response to the magnetic fields (see Scheme 1(a) & (b)). One is the intrinsic Néel motion (rotating magnetic moment inside the stationary MNP) and the other is the extrinsic Brownian motion (rotating the entire MNP along with its magnetic moment).

The relaxation dynamics of MNPs are dependent on the balance between the magnetic forces that tend to align the magnetic moments of MNPs along the external magnetic field direction and the thermal fluctuations that hinder the alignments. Relaxation time models are used to characterize these mechanisms.

Zero-field Brownian relaxation time is expressed as:[1]

$$\tau_B|_{zero-field} = \tau_{B0} = \frac{3\eta V_h}{k_B T} \quad (1),$$

where $\eta$ is the viscosity of ferrofluid, $k_B$ is Boltzmann constant, $T$ is temperature, and $V_h$ is the hydrodynamic volume of nanoparticles. In this paper, iron oxide nanoparticles with magnetic core diameters $D$ and surface chemical compounds thickness of $c$ are assumed. So, $V_h = \pi(D + 2c)^3/6$.

Zero-field Néel relaxation time is expressed as:[2]

$$\tau_N|_{zero-field} = \frac{\sqrt{\pi}}{2}\tau_{N0}\sigma^{-3/2}e^{\sigma} \quad (2),$$

where,

$$\tau_{N0} = \frac{\beta(1+\alpha'^2)M_s}{2\gamma\alpha'} \quad (3)$$

$$\sigma = \frac{-\Delta E}{k_B T} \quad (4)$$

$$\beta = \frac{V_c}{k_B T} \quad (5),$$

where $\gamma$ is electron gyromagnetic ratio ($1.76 \times 10^{11}\ rad/sT$), $\alpha'$ is damping constant, which is typically on the order of $10^{-3}$~$10^{-2}$ for bulk materials.[3] While fine particles have larger damping constants than bulk, usually $\alpha' = 0.1$ for iron oxide nanoparticles.[2] $\Delta E$ is the energy barrier a nanoparticle needs to overcome before flipping from one minimum energy state to the other minimum energy state. Magnetic core volume $V_c = \pi D^3/6$.

For MPS measurements, external magnetic fields are applied to the MNP ferrofluid systems. Hence, the zero-field relaxation time equations aforementioned do not apply in this case. These equations only give us the time for net magnetization to relax back to equilibrium after the removal of external magnetic fields.[2] As is reported before, the relaxation time is not only dependent on the externally applied magnetic field strength but also dependent on the field frequency.[4]

Analytical expression for Brownian relaxation time at nonzero field is:[5]

$$\tau_B = \tau_{B0}\frac{1+\alpha^2-\alpha^2\coth^2\alpha}{\alpha\coth\alpha-1} \quad (6)$$

Analytical expression for Néel relaxation time at nonzero field is:[6]

$$\tau_N = \frac{\tau_{N0}}{\sigma_{eff}(1-h^2)} \left( \frac{\sqrt{\sigma_{eff}/\pi}}{1+1/\sigma_{eff}} + 2^{-\sigma-1} \right)^{-1} \times \left( \frac{1-h}{e^{\sigma_{eff}(1-h)^2}-1} + \frac{1+h}{e^{\sigma_{eff}(1+h)^2}-1} \right)^{-1} (7),$$

where $\alpha$ and $h$ are time-varying parameters modulated by external magnetic field $H(t)$:

$$\alpha(t) = \frac{M_s V_c}{k_B T} \mu_0 H(t) \quad (8)$$

$$h(t) = \frac{\alpha(t)}{2\sigma_{eff}} \quad (9)$$

As we apply a periodically varying external magnetic field $H(t)$ to the ferrofluid system, the relaxation time $\tau_B$ and $\tau_N$ will also show a periodical change synchronously.

The magnetization dynamics of MNPs are usually characterized by effective relaxation time $\tau_{eff}$, which is dependent on Brownian relaxation time $\tau_B$ and Néel relaxation time $\tau_N$. Both relaxation processes are dependent on the frequency and amplitude of applied magnetic fields. The $\tau_{eff}$ of a nanoparticle governs its ability to follow the external fields. The effective relaxation time $\tau_{eff}$ is related to the Brownian and Néel relaxation times:

$$\frac{1}{\tau_{eff}} = \frac{1}{\tau_B} + \frac{1}{\tau_N} \quad (10)$$

$\tau_{eff}$ of a nanoparticle governs its ability to follow the alternating field.

**Note S3. Langevin Model of Magnetic Responses**

In the presence of oscillating magnetic fields, MNPs are magnetized and their magnetic moments tend to align with the fields. For a ferrofluid system of monodispersed, noninteracting MNPs, the magnetic response obeys Langevin function:

$$M_D(t) = m_s c L(\xi) \quad (11),$$

where,

$$L(\xi) = \coth \xi - \frac{1}{\xi} \quad (12)$$

$$\xi = \frac{m_s H(t)}{k_B T} \quad (13)$$

The MNPs are characterized by magnetic core diameter $D$, saturation magnetization $M_s$ and concentration $c$. Assuming MNPs are spherical without mutual interactions. The magnetic moment of each particle is $m_s = M_s V_c = M_s \pi D^3/6$, where $V_c$ is volume of the magnetic core, $\xi$ is the ratio of magnetic energy over thermal energy, $k_B$ is Boltzmann constant, and $T$ is the absolute temperature in Kelvin. $H(t) = A_H \cos(2\pi f_H t) + A_L \cos(2\pi f_L t)$ are the external magnetic fields, where $A_H$, $A_L$, $f_H$, $f_L$ are the amplitude and frequency of high and low frequency driving fields, respectively.

It should be noted that the Langevin model is not able to describe the phase information of the harmonic signals. When the frequency of the driving magnetic field is low, the SPION's magnetic moment can follow the driving field tightly and the magnetic susceptibility $\chi$ is a real number. As the driving field frequency increases, a phase lag $\varphi$ between the SPION's magnetic moment and the driving magnetic field is introduced, which leads to a complex magnetic susceptibility $\chi$ and the Debye model is used in this case.

**Note S4. The Phase Lag Model**

The phase lag is modulated by the low-frequency field and also can be monitored at the harmonic phase angles:

$$\varphi(t) = arctan(\omega\tau_{eff}) \quad (15)$$

where $\omega$ is the angular frequency, $\tau_{eff}$ is modulated by the oscillating magnetic field.

## Note S5. The Induced Voltage in Pick-up Coils

According to Faraday's law of Induction, the induced voltage in a pair of pick-up coils is expressed as:

$$u(t) = -S_0 V \frac{d}{dt} M_D(t) \quad (14)$$

where $V$ is volume of MNP suspension. Pick-up coil sensitivity $S_0$ equals to the external magnetic field strength divided by current.

## Note S6. The 3rd and the 5th Harmonics

Taylor expansion of $M_D(t)$ from equation (11) shows the major frequency mixing components:

$$\frac{M_D(t)}{m_s c} = L\left(\frac{m_s H(t)}{k_B T}\right)$$

$$= \frac{1}{3}\left(\frac{m_s}{k_B T}\right) H(t) - \frac{1}{45}\left(\frac{m_s}{k_B T}\right)^3 H(t)^3 + \frac{2}{945}\left(\frac{m_s}{k_B T}\right)^5 H(t)^5 + \cdots$$

$$= \cdots + \left[-\frac{1}{60} A_H A_L^2 \left(\frac{m_s}{k_B T}\right)^3 + \cdots\right] \times \cos[2\pi(f_H \pm 2f_L)t + \varphi_{f_H \pm 2f_L}]$$

$$+ \left[\frac{1}{1512} A_H A_L^4 \left(\frac{m_s}{k_B T}\right)^5 + \cdots\right] \times \cos[2\pi(f_H \pm 4f_L)t + \varphi_{f_H \pm 4f_L}]$$

$$+ \cdots$$

(16)

The mixing frequency components are found at odd harmonics exclusively:

$$M_D(t)|_{3rd} \approx \frac{-m_s c}{60} A_H A_L^2 \left(\frac{m_s}{k_B T}\right)^3 \times \cos[2\pi(f_H + 2f_L)t + \varphi_{f_H \pm 2f_L}] \quad (17)$$

$$M_D(t)|_{5th} \approx \frac{m_s c}{1512} A_H A_L^4 \left(\frac{m_s}{k_B T}\right)^5 \times \cos[2\pi(f_H + 4f_L)t + \varphi_{f_H \pm 4f_L}] \quad (18)$$

Amplitudes of induced voltages at the 3rd and 5th harmonics are expressed as:

$$u_{3rd} = -S_0 V \frac{d}{dt} M_D(t)|_{3rd} \quad (19)$$

$$u_{5th} = -S_0 V \frac{d}{dt} M_D(t)|_{5th} \quad (20)$$

**Note S7. The 3rd over the 5th Harmonic Ratio (R35) as MNP Quantity-Independent Metric**

Harmonic ratio of the 3rd over the 5th harmonics is expressed as:

$$R35 = \frac{u_{3rd}}{u_{5th}} = \frac{S_0 V \frac{d}{dt} M_D(t)|_{3rd}}{S_0 V \frac{d}{dt} M_D(t)|_{5th}} = \frac{\frac{d}{dt} M_D(t)|_{3rd}}{\frac{d}{dt} M_D(t)|_{5th}} \quad (21)$$

Hence, one advantage of using this harmonic ratio to characterize magnetic properties of MNPs is that this parameter is independent of concentration/quantity of MNPs in the testing sample.

**Note S8. The R35 Heatmap**

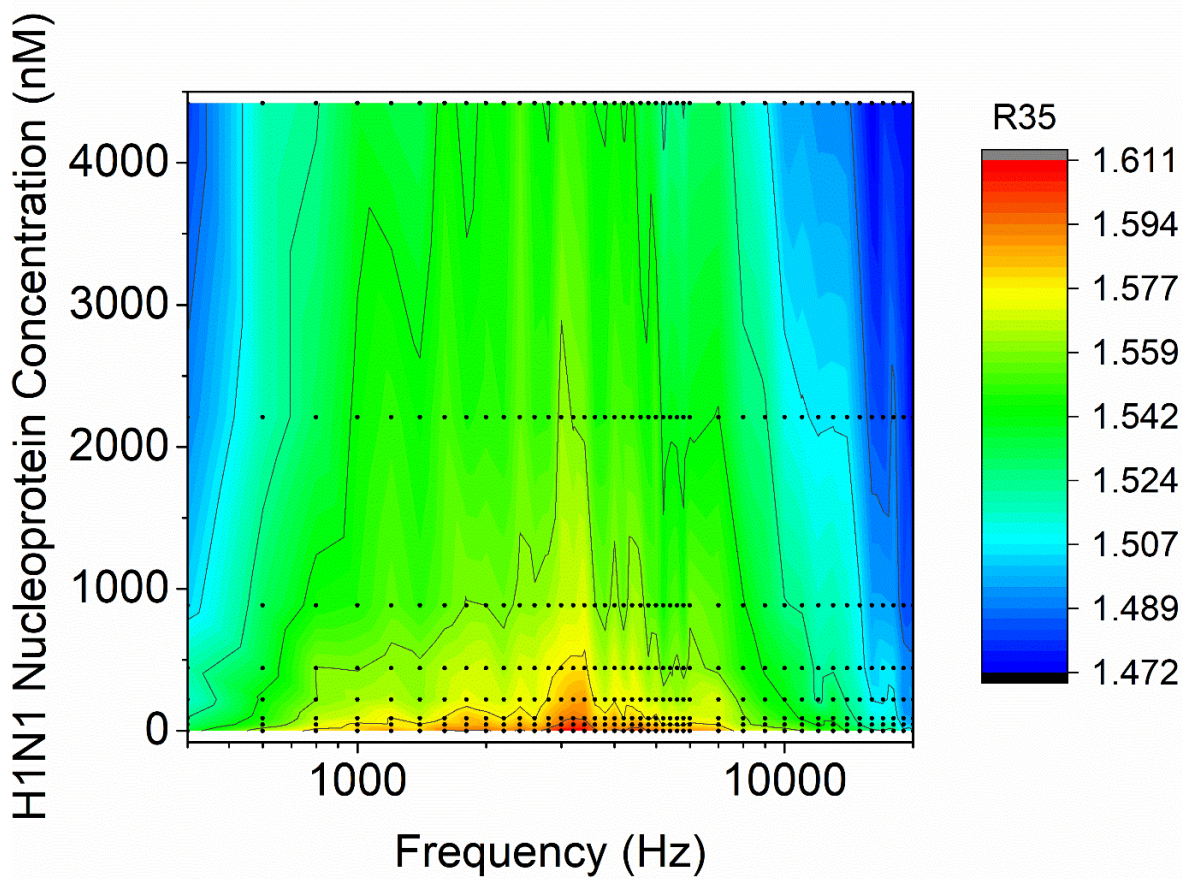

Figure S2. The heatmap of R35 values at combinational H1N1 nucleoprotein concentrations and driving field frequencies.

**Note S9. Capabilities of the 3rd Harmonic, the 5th Harmonic, and the Harmonic Ratio R35 as Metrics for Distinguishing H1N1 Nucleoprotein Samples with High and Low Concentrations.**

It should be noted that both the 3rd and 5th harmonics show significantly weaker capabilities in detecting low concentrations of H1N1 nucleoprotein (see the red curves in Figure S3(a) & (b)). Under the driving field frequency of 1 kHz, the differences in 3rd harmonic amplitudes measured from samples VII (44 nM) and VIII (0 nM) is only 2.8%, this difference increases to 3.9% at a driving field frequency of 5 kHz (see Figure S3(a)). Similarly, the differences in 5th harmonic amplitudes from samples VII (44 nM) and VIII (0 nM) is only 2.6% at 1 KHz and increases to 3.1% at 5 kHz (see Figure S3(b)). However, neither the 3rd harmonic nor the 5th harmonic succeed in distinguishing samples I (4.42 µM) and II (2.21 µM) where the concentrations of H1N1 nucleoprotein are extremely high, at 20 kHz.

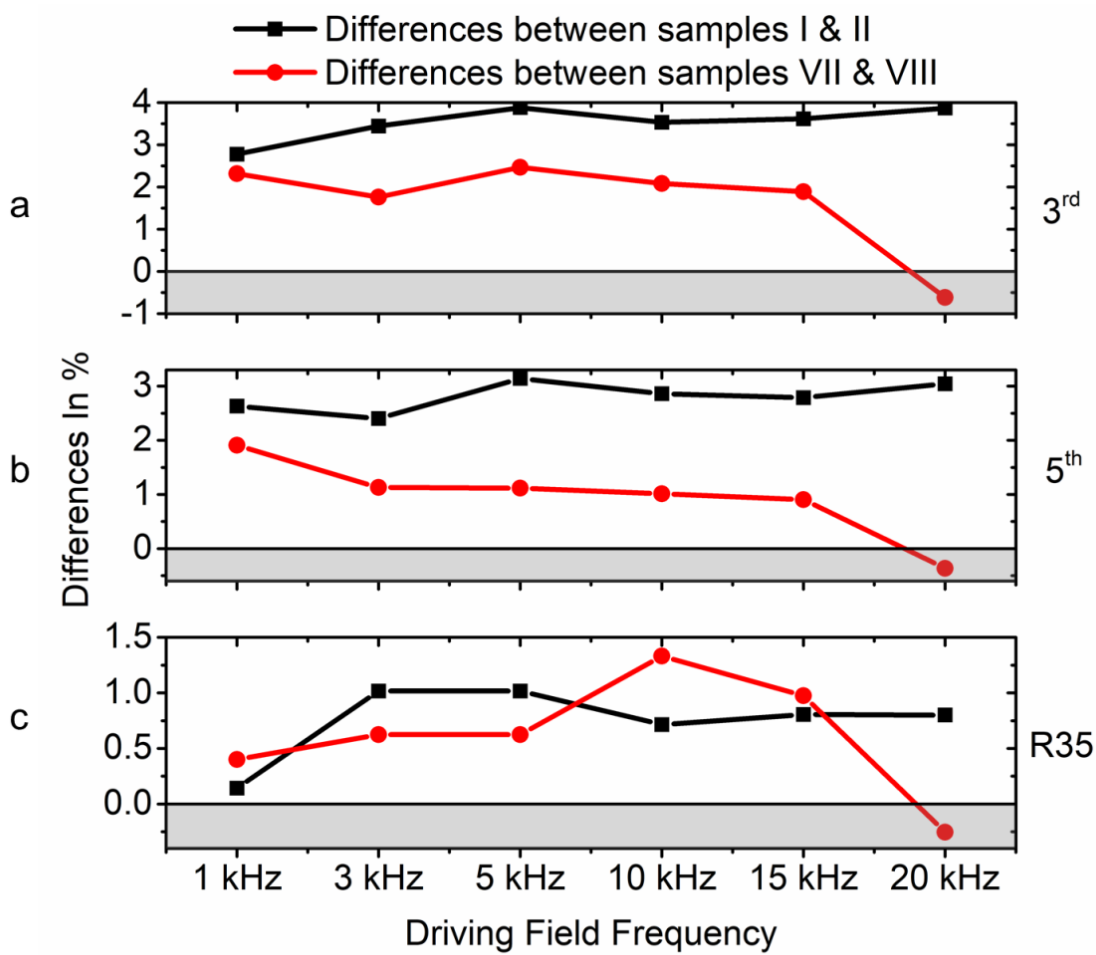

Figure S3. (a) The 3rd harmonic, (b) the 5th harmonic, and (c) the harmonic ratio R35 as metrics for distinguishing samples with low and high H1N1 nucleoprotein concentrations. The grey area indicates that the metric fails to differentiate samples VII & VIII at 20 kHz.

On the other hand, the harmonic ratio R35 demonstrates comparable capabilities in distinguishing high and low concentrations of H1N1 nucleoprotein samples (see Figure S3(c)). For the detection of samples with low concentrations of H1N1 nucleoproteins, the R35 differences from samples VII (44 nM) and VIII (0 nM) increase from 0.14% to 1.02% from 1 kHz to 5 kHz. At the higher end of H1N1 nucleoprotein concentrations, the harmonic ratio R35 shows stronger capability in distinguishing samples I (4.42 µM) and II (2.21 µM) with differences of 1.33% and 0.97% at 10 kHz and 15 kHz, respectively (see the red curve in Figure S3(c)).

Table S1. The 3rd harmonic amplitudes (µV) from Figure 2(a):i – iv.

| Sample Index | I | II | III | IV | V | VI | VII | VIII | IX |
|---|---|---|---|---|---|---|---|---|---|
| **1 kHz** | 138 | 141.2 | 148.4 | 150.75 | 162.6 | 168.6 | 180 | 185 | 184.8 |
| **5 kHz** | 219 | 224.4 | 236.5 | 241.66 | 262 | 272.75 | 290 | 301.25 | 306.25 |
| **10 kHz** | 199.25 | 203.4 | 215 | 219.8 | 238 | 246.75 | 263.2 | 272.5 | 280.25 |
| **20 kHz** | 162.4 | 161.4 | 171.4 | 173.6 | 188 | 193.8 | 206.8 | 214.8 | 219 |

Table S2. The 5th harmonic amplitudes (µV) from Figure 2(b):v – viii.

| Sample Index | I | II | III | IV | V | VI | VII | VIII | IX |
|---|---|---|---|---|---|---|---|---|---|
| **1 kHz** | 89.728 | 91.441 | 95.878 | 96.711 | 104 | 107.4 | 114 | 116 | 117 |
| **5 kHz** | 143 | 144.6 | 152.25 | 154.33 | 167 | 172.75 | 183 | 188.75 | 190 |
| **10 kHz** | 133.25 | 134.6 | 141 | 143 | 154.4 | 159.5 | 169.4 | 174.25 | 177.5 |
| **20 kHz** | 110 | 109.6 | 115.4 | 116.4 | 125.6 | 129.6 | 138 | 142.2 | 143.8 |